\documentclass[twocolumn,showpacs,amsmath,amssymb,prl,superscriptaddress,floatfix]{revtex4}

\usepackage{graphicx}
\usepackage{dcolumn}
\usepackage{bm}

\usepackage{amssymb}
\usepackage{amsmath}
\usepackage{amsfonts}
\usepackage{amssymb}
\usepackage{hyperref}
\usepackage{color}

\usepackage{xcolor}
\definecolor{red}{rgb}{0.7,0,0}
\definecolor{green}{rgb}{0.,0.35,0.}
\definecolor{blue}{rgb}{0.2,0.2,0.7} 
\definecolor{black}{rgb}{0.15,0.15,.15}

\begin{document}

\title{1D Quantum Liquids with Power-Law Interactions: the Luttinger Staircase}

\date{\today}
\author{M. Dalmonte}
\affiliation{Dipartimento di Fisica dell'Universit\`a di Bologna and INFN, via Irnerio 46, 40127 Bologna, Italy}
\affiliation{IQOQI and Institute
for Theoretical Physics, University of Innsbruck, A-6020 Innsbruck, Austria}
\author{G. Pupillo}
\affiliation{IQOQI and Institute
for Theoretical Physics, University of Innsbruck, A-6020 Innsbruck, Austria}

\author{P. Zoller}
\affiliation{IQOQI and Institute
for Theoretical Physics, University of Innsbruck, A-6020 Innsbruck, Austria}
\affiliation{Norman Bridge Laboratory of Physics 12-33, California Institute of Technology, Pasadena, CA 91125 USA}

\begin{abstract}
We study one dimensional fermionic and bosonic gases with repulsive power-law interactions $1/|x|^{\beta}$, with $\beta>1$, in the framework of Tomonaga-Luttinger liquid (TLL) theory. We obtain an accurate analytical expression linking the TLL parameter to the microscopic Hamiltonian, for arbitrary $\beta$ and strength of the interactions. In the presence of a small periodic potential, power-law interactions make the TLL unstable towards the formation of a cascade of lattice solids with fractional filling, a ``Luttinger staircase". Several of these quantum phases and phase transitions are realized with groundstate polar molecules and weakly-bound magnetic Feshbach molecules.
\end{abstract}
\maketitle

There is presently considerable interest in quantum degenerate gases with long range interactions in reduced geometries~\cite{LewensteinReview}. This is motivated by recent experiments with polar molecules~\cite{PolMolExp}, where electric dipole moments associated with rotational excitations lead to strong, anisotropic dipolar interactions~\cite{Ni}, but also by experiments with atomic gases with strong magnetic dipoles~\cite{pfau}. For polar molecules, electric dipoles can be manipulated with external microwave AC and DC electric fields, which provides a toolbox to tailor the many-body interactions, and in combination with optical trapping in 1D or 2D promises the realization of stable exotic, strongly correlated quantum phases with long range interactions~\cite{shape}.

\begin{figure}[b]{
\includegraphics[width= 0.93\columnwidth]{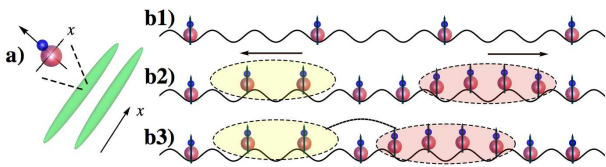}
\caption{\small{(color online)(a) Experimental setup (sketch): an array of 1D polar molecular gases  is formed along $x$ (green tubes); 
molecules are polarized perpendicular to $x$.} (b1) Groundstate configuration in the solid phase with filling $1/p=1/3$. (b2) soliton and antisoliton excitations with repulsive interactions with $1/p=1$. (b3) a breather, $1/p=1$ (text).}
 \label{fig:fig1}}
\end{figure}

An intriguing  example is given by polar molecules trapped in a 1D wire [see Fig.~\ref{fig:fig1}(a)]~\cite{Kollath,citro,citro2,roscilde}, where long range interactions compete with an optical lattice in a commensurate - incommensurate transition. In the zero-tunneling limit in a deep lattice, this leads to the formation of a {\em devil's staircase}, that is, a continuous and non differentiable (Cantor) function for the ground state filling fraction as function of the chemical potential $\mu$, studied first in the context of atomic monolayers adsorbed on solids~\cite{devil}. While recent studies at finite hopping have already shown a modification of this structure in a deep lattice~\cite{sondhi}, the challenge is now to investigate the quantum regime where large kinetic energies compete with both interaction strengths and periodic confinement.

Below we show that using bosonization techniques~\cite{bosonization} the classification of quantum phases can be derived analytically for all power law interactions $C_\beta /|x|^\beta$ with $\beta >1$ and for arbitrary relative strengths of the kinetic energy and the long-range repulsion. Remarkably, the parameters of the effective bosonized theory can be accurately obtained in analytical form for all $\beta$ in terms of the microscopic Hamiltonian, even in the absence of integrability. This provides us with a universal phase diagram where the cases of repulsive Van der Waals ($\beta=6$) and dipolar interactions ($\beta=3$) should be accessible in polar molecule experiments~\cite{shape}. In contrast to the classical devil's staircase, where lattice solids are stable over a finite interval in $\mu$ for every rational filling fraction between $0$ and $1$, and the total measure of such interval exhausts the full range of $\mu$, we find that in the 1D quantum case large kinetic energies prevent the formation of ordered states, where the average interparticle distance is not constant. This drastically reduces the number of "steps" in the staircase to a number not dense in the interval $]0,1]$, i.e. a {\em Luttinger staircase}. Signatures of these quantum phases are excitations in the form of solitons and breathers, detectable via Bragg scattering.

We assume that the polar molecules are polarized by external electric fields, and confined to a 1D geometry, e.g., by a sufficiently deep 2D optical lattice with frequency $\omega_\perp$. The shape of the long distance interactions can be tuned by coupling the lowest rotational manifolds of each molecule with DC and microwave AC fields. As shown in Refs.~\cite{shape} we can tune between $\beta =3$  with $C_3 = d^2/\epsilon_0$, where $d$ is the dipole moment induced by an electric field $E_{\rm DC}$ and $\epsilon_0$ the vacuum permittivity, and  $\beta =6$ with $C_6 \propto d^4/\hbar \Delta$, where $\Delta$ the detuning of a microwave field $E_{\rm AC}$ coupling the ground to the first excited rotational manifold. For average interparticle distances $a \gg (C_\beta/\hbar \omega_\perp)^{1/\beta}$ the gas dynamics is one-dimensional and microscopically described by the Hamiltonian
\begin{eqnarray}
H &=& \int dx \: \psi^{\dagger}(x)\left[-\frac{\hbar^2 }{2m}\partial_x^2+U(x)\right]\psi(x) \nonumber\\
 &+&  \frac{1}{8 \pi} \int dx \:dx' \psi^{\dagger}(x)\psi^{\dagger}(x')\frac{C_{\beta}}{|x-x'|^{\beta}}\psi(x')\psi(x) .\label{eq:eqHam}
\end{eqnarray}
Here, $\psi(x)$ is a field operator for molecules, which can be either fermionic or bosonic, $m$ is the particle mass, and $U(x)=U \sin^2 (2\pi x/\lambda )$ is a weak periodic potential, as provided by a shallow optical lattice of strength $U_L\equiv U/E_R \lesssim 1$, with $E_R=h^2/2m\lambda^2$ and $\lambda$ the lattice wavelength.
For the case $\beta=3$ with experimentally relevant molecules such as LiCs, RbCs or KRb molecules ($d_{\max}=5.6$, 1.25 and 0.5 Debye, respectively) and confinement $\omega_\perp = 2 \pi \times 100$ kHz, $(C_\beta/\hbar \omega_\perp)^{1/3}$ is of the order of 360, 130 and 80nm, respectively~\cite{shape,micheli10}.

{\it In the absence of an optical lattice} ($U_L=0$) the short range character of power-law interactions with $\beta>1$ allows a description of the low energy physics in terms of Tomonaga-Luttinger liquid (TLL) theory~\cite{bosonization,schulz}.  Here, we first consider the bosonic case, and then discuss the differences with the fermionic one. The TLL effective Hamiltonian is given by~\cite{haldane,citro}
\begin{equation}
H=\hbar v\int dx\left[(\partial_{x}\theta(x))^{2}/K+K(\partial_{x}\phi(x))^{2}\right]/(2\pi).\label{eq:eqLL}\end{equation}
Here, the field $\psi(x)$ in Eq.~\eqref{eq:eqHam} is replaced
by $\psi(x)\sim\sqrt{n+\partial_{x}\theta(x)/\pi}\exp[i\phi(x)]$
in a hydrodynamic approach, and $\partial_{x}\theta(x)$ and $\partial_{x}\phi(x)$
characterize the long-wavelength fluctuations of the density $n$
and of the phase $\phi(x)$, respectively, with $[\partial_{x}\theta(x),\phi(y)]=i\pi\delta(x-y)$.
The liquid is completely characterized by the sound velocity $v$
and Luttinger parameter $K=\hbar n\pi/(mv)$, which determines the
algebraic decay of the correlation functions
\begin{equation}
\langle n(x)n(x')\rangle\sim|x-x'|^{-2K},\;\langle\psi(x)\psi^{\dagger}(x')\rangle\sim|x-x'|^{-1/2K}.\label{eq:corr}\end{equation}

In general, $K$ can be related to the microscopic parameters of the Hamiltonian
only for exactly solvable models, e.g., contact interactions or $\beta=2$ [Calogero-Sutherland (CS) model].
Below we show that the dependence of $K$ on the microscopic parameters in~\eqref{eq:eqHam}
can be given \emph{analytically} for arbitrary shape and strength of interactions,
\begin{equation}
K=\left[1+\beta(\beta+1)\zeta(\beta)n^{\beta-2}R_{\beta}/(2\pi^{2})\right]^{-1/2},\label{eq:eqK}\end{equation}
with $n^{\beta-2}R_{\beta}$ the dimensionless interaction strength,
and $R_{\beta}\equiv mC_{\beta}/(2\pi\hbar^{2})$ (see Fig.\ref{fig:fig3}). 
In contrast to familiar bosonic gases
with contact interactions where $K\geq1$~\cite{bosonization,cazalilla,buchler},
long-range power-law interactions constrain $K$ to values $1\ge K>0$, where $K=1$ corresponds to the Tonks-Girardeau gas limit and $K=0$ to a system with long-range order~\cite{schulz}. Eq.~\eqref{eq:eqK} allows us to readily determine the \emph{phase diagram} for $U_{L}=0$, by comparing the relative decay of the correlation functions in~\eqref{eq:corr}: a crossover from superfluid (SF) to charge-density wave (CDW) behavior takes place at $K_{c}=0.5$. The fermionic gas is also described by Eqs.~\eqref{eq:eqLL} and~\eqref{eq:eqK}, however its phase diagram displays a CDW behavior at all interaction strengths. In addition, correlation functions in~\eqref{eq:corr} have a slightly different long-distance decay~\cite{bosonization}. In the following, statistics will not be relevant, and thus we deal with both cases at the same time.

\begin{figure}[t!]
\center{{\includegraphics[width= 0.9 \columnwidth]{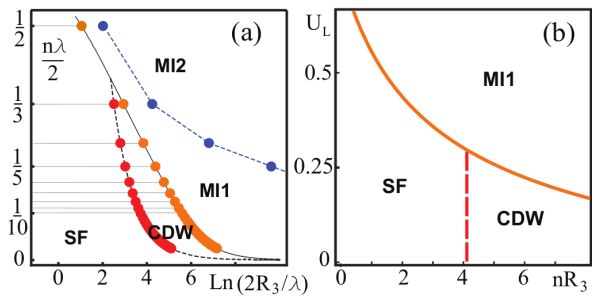}}
\caption{(color online) (a) Commensurate phase diagram for bosons with dipolar interactions $\beta=3$, and lattice depth $U_L = 0.1$. Physical configurations correspond to commensurate fillings $n \lambda/2 \equiv 1/p$, with $p \in \mathbb{N}$ (horizontal lines are guides to the eye for $p \leq 10$). Quantum phase transitions from a TLL to a lattice solid [or Mott insulator, (MI)] occur for each $1/p$ at the position of the dots on the continuous line, while red and blue dots on dashed lines indicate crossovers. MI1 and MI2 indicate MI with solitonic and breather excitations, respectively (see Fig.1). (b) Phase diagram at commensurate filling $1/p= 1/3$ in the $U_L$ vs $nR_3$ plane. Continuous line: quantum phase transition between a TLL and a lattice solid. The phase diagram for fermions is identical to the one for bosons, except the TLL is always a CDW.}
\label{fig:fig2}}
\end{figure}

{\it The Luttinger staircase:} A cascade of insulating lattice solids can be realized from a TLL with power-law interactions, by introducing a vanishingly-small periodic lattice potential, as provided by the shallow optical lattice $U(x)$ in Eq.~\eqref{eq:eqHam}. Combining the complete density operator with the periodic term $U(x)$~\cite{buchler}, one obtains
\begin{equation}\label{eq:eqsG}
\sum_r U_r\equiv \sum_r \left\{ \frac{\mathcal{U}}{\pi\Lambda^2}\int \cos[r2\theta(x)+Q_{r}x] dx\right\},
\end{equation}
with $Q_{r}(n,\lambda)\equiv 2\pi \left(rn-2/\lambda\right)$ and $r\in\mathbb{N}$. Here, $\mathcal{U} \equiv \pi n U \Lambda^2/2$, where $\Lambda$ is a cutoff that fixes the energy scale of the initial Hamiltonian~\cite{bosonization}. The term $Q_r$ in Eq.~\eqref{eq:eqsG} is responsible for a possible competition between two lengths scales: the interparticle distance $1/n$ and the lattice period $\lambda/2$. We can then distinguish two different situations: a \emph{commensurate} one with $Q_r=0$, where the length-scales do not compete, corresponding to the condition $2/(\lambda n)=p \in \mathbb{N}$, and an \emph{incommensurate} one with $Q_r\neq 0$, where a competition is present.

In all commensurate cases  $2/(\lambda n)=p\in \mathbb{N}$, the most relevant term due to the optical lattice in~\eqref{eq:eqsG} in the renormalization-group sense is $U_p$. Keeping only this term, the system becomes equivalent to a sine-Gordon model~\cite{bosonization}, where the scaling dimension of the cosine operator is affected by the interparticle interactions through $K$ and by the optical lattice through $\mathcal{U}$. For $p>1$ we then expect that for weak interactions and small depth of the optical lattice the cosine term in~\eqref{eq:eqsG} is irrelevant and the TLL liquid is preserved, with correlation functions decaying polynomially as in~\eqref{eq:corr}. However, when the cosine term is relevant, we have a non-zero expectation value  $\langle\cos[p \sqrt{4\pi}\phi(x)]\rangle\neq 0$ and  the system is \emph{pinned} on the lattice. This pinning corresponds to the breakdown of TLL and to the formation of a lattice solid, or Mott insulator (MI), with particles localized at individual sites of the lattice, every $p$ lattice sites. In this phase, the excitation spectrum is gapped and the off-diagonal correlation function decays exponentially. Using Berezinskii-Kosterlitz-Thouless (BKT) scaling near criticality~\cite{bosonization}, we find that for each $p$ the gapped phase occurs for
\begin{equation}\label{eq:eqBKT}
2 + U_L>  K p^2,
\end{equation}
with $U_L \equiv U/E_R \lesssim 1$. Eq.~\eqref{eq:eqBKT} is remarkable in that it shows that power-law interactions make possible the realization of an infinite series of gapped phases at lattice filling less than one [Fig.~\ref{fig:fig1}(b1) for $p=3$]. The case $p=1$ is peculiar as Eq.~\eqref{eq:eqBKT} is always satisfied, implying a lattice solid for a vanishingly small $U_L$. The cascade of solids with $p \in \mathbb{N}$ corresponds to a quantum version of a Devil's staircase structure, where large kinetic energies of order of $E_R$ prevent the formation of ordered states where the average interparticle distance is not constant. This is in contrast to the {\it classical} Devil's staircase of the Frenkel-Kontorova model~\cite{devil}, where commensurability is also allowed for rational fillings $r \in \mathbb{Q} \neq \mathbb{N}$. Evidence of this classical case have been recently found in the deep lattice limit of Refs.~\cite{sondhi} for $\beta=3$, in 1D and 2D.

Equation~\eqref{eq:eqBKT} shows that a gap is favored by high densities, strong interactions and finite (small) lattice depths. For $U_L<1$ a good estimate for the gap $\Delta$ is given by~\cite{zam}
\begin{equation}
\Delta=\frac{8}{\sqrt{\pi}}\frac{\Gamma[\frac{\pi K}{(4-2K)}]}{\; \Gamma[\frac{2+K(\pi-1)}{4-2K}]}\left(K^2\frac{U_L}{16}\frac{\Gamma[1-\frac{K}{2}]}{\Gamma[1+\frac{K}{2}]}
\right)^{\frac{1}{2-K}}.
\end{equation}
When $K$ is close to $1/p^2$, $\Delta$ approaches the \emph{massive fermion} limit $\Delta\sim U/2$ recently observed for $p=1$ with contact interactions~\cite{buchler,Haller2010}, whereas close to the BKT transition it closes exponentially. In the vicinity of the BKT transition excitations are of the soliton/antisoliton type, which in the massive fermion limit correspond to weakly repulsive particles and holes, Fig.~\ref{fig:fig1}(b2). In contrast to contact interactions, power-law interactions allow one to tune the {\it sign} of the soliton-antisoliton interactions from repulsive ($K>1/p^2$) to {\it attractive} ($K<1/p^2$), giving rise to soliton-antisoliton bound states called \emph{breathers} [Fig.~\ref{fig:fig1}(b3)]. These excitations are confined in space but oscillatory in time, and are stable solutions of the equations of motion for the sine-Gordon model~\cite{bosonization}. The number of breather excitations is $\mathcal{N}=2(1/K-1)$, with energy
\begin{equation}
M_b(n')=2\Delta \sin[\pi n'/(4/K-2)], \quad n' \leq \mathcal{N}.
\end{equation}
For $K<1/(2p^2)$ breathers are the lowest-energy excitations, qualitatively changing the spectrum of the insulating phase with respect to the familiar case of contact interactions. Strong power-law interactions will allow for an unambiguous observation of this localized topological excitations, with applications ranging from Josephson junctions to conjugated polymers, see below~\cite{breathers}.
\begin{table}[t!]
\begin{tabular}{c|c|c|c|c|c|c}
   & KRb & RbCs & LiCs & $^{52} {\rm Cr}_2$ &  $^{164}{\rm Dy}_2 $ & $^{166}{\rm Er}_2$ \\
   \hline
  \hline
  $U_L(1)$ &  $0_+$ & $0_+$ & $0_+$ & $0_+$ & $0_+$ & $0_+$ \\
  \hline
  $U_L(1/2)$ &  0.8 & $0_+$ & $0_+$ & $1.9$ & $1.5$ & $1.7$\\
  \hline
  $U_L(1/3)$ & - & 1.4 & $0_+$ & - & - & - \\
\end{tabular}
\caption{Minimal lattice depth $U_L(1/p)$ for a lattice solid at filling $1/p$, for groundstate polar molecules (KRb, RbCs, LiCs) and magnetic Feshbach molecules ($^{52} {\rm Cr}_2$,$^{164}{\rm Dy}_2 $,$^{166}{\rm Er}_2$); an arbitrarily small periodic potential pins the TLL for $0_+$. Lattice depths with $U_L(1/p) \gtrsim 2$, where the sine-Gordon picture breaks down~\cite{Haller2010}, are not considered.}\label{table:table1}
\end{table}
\begin{figure}[b]{
\includegraphics[width= 0.8\columnwidth]{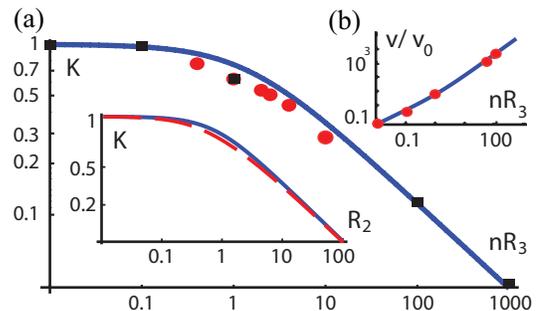}
\caption{\small{(color online) (a) TLL parameter $K$ vs the dimensionless interaction strength $n^{\beta-2} R_\beta$ for dipolar interactions $\beta=3$. Line: analytic result Eq.~\eqref{eq:eqK}. Squares and dots: quantum Monte-Carlo (QMC) results of Refs.~\cite{citro} and~\cite{roscilde}, respectively. Inset: $K$ vs $n^{\beta-2} R_\beta$ with $\beta=2$. Continuous line: Eq.~\eqref{eq:eqK}. Dashed line: exact  Calogero-Sutherland model. (b) TLL velocity $v$ vs $n R_3$ for $\beta =3$, with $v_0 \equiv \hbar/(\sqrt{2}mR_3)$. Line and dots: analytic and QMC results of Ref.~\cite{citro2}, respectively.}}
 \label{fig:fig3}}
\end{figure}
In the vicinity of commensurate fillings ($p \in \mathbb{N}$), the system exhibits a gap as long as the energy shift due to $Q_r\neq 0$ remains small with respect to $K \Delta/2$, the energy required to add a particle: this ensures stability of the phases above with respect to small density changes. Above a critical $Q_c$ a commensurate-incommensurate phase transition takes place from an insulator to a gapless phase, similar to contact interactions~\cite{buchler}. For generic values of $Q_r\neq 0$, the TLL is stable.

Figure~\ref{fig:fig2}(a) shows the {\it commensurate phase diagram} for the case of bosonic particles with dipole-dipole interactions $\beta=3$ as a function of the lattice filling $1/p = n \lambda/2$ and the strength of interactions $R_3$. The lattice depth is $U_L = 0.1$. For each $p \in \mathbb{N}$, the BKT quantum phase transition occurs at the position of the dot along the continuous line, while dots on the dashed lines characterize crossovers. The regions denoted as MI1 and MI2 correspond to MI with soliton/antisoliton and breather excitations, respectively, and the dashed line signals the crossover for $K=1/(2p^2)$. Panel (b) shows the transition between the TLL and solid behavior as a function of the lattice depth and $n R_3$ for the case of $p=3$. The phase diagram for fermionic particles is identical to Fig.~\ref{fig:fig2}, except that the TLL phase is always a CDW. Phase diagrams for $\beta\neq 3$ look qualitatively similar to Fig.~\ref{fig:fig2}.

In Table~\ref{table:table1} we list the estimated minimal lattice depth necessary to realize a MI with filling $1/p$, for a few groundstate polar molecules.
In addition, we report estimates for magnetic Feshbach molecules,
where the magnetic dipole moment is taken as twice the atomic one~\cite{Ferlaino09}.
The realization of insulating states with, e.g., $p=2$ will help stabilize highly-excited Feshbach molecules against three-body recombination, opening the way to the realization of strongly-correlated lattice phases.

\emph{Analytical expression for $K$}: 
For $U_L=0$, Eq.~\emph{\eqref{eq:eqHam}} describes an effective, strictly one-dimensional,
{\it scale invariant} theory, dependent only on $n^{\beta-2}R_{\beta}$ at all length/energy scales.
After rescaling, dimensionless interactions read $V(y\equiv rn)=R_\beta n^{\beta-2}/y^{\beta}$, and ultraviolet divergences in Eq.~\emph{\eqref{eq:eqHam}} can be treated by introducing a dimensionless cut-off $A$~\cite{bosonization}. We choose $A$ such that $V(y)\propto (y+A)^{-\beta}$, so that scale-invariance is preserved explicitly: the {\it shape} of the regularized potential is independent of $C_{\beta}$ and $n$. We can now fix $A$ self-consistently in the effective 1D theory, as follows [consistency with the microscopic derivation of~\emph{\eqref{eq:eqHam}} implies $A/n \gtrsim (C_\beta/\hbar \omega_\perp)^{1/\beta}$]. We compute $K$ analytically in the strong- ($n^{\beta-2}R_{\beta}\gg 1$) and in the weak-coupling ($n^{\beta-2}R_{\beta}\ll 1$) limits as
$K_{\rm s}=\pi/[\beta(\beta+1)\zeta(\beta)R_{\beta}n^{\beta-2}/2]^{1/2}$~\cite{citro2} and $K_{\rm w}=(1+n^{\beta-2}R_{\beta} \tilde{V}(0)/2\pi)^{-1/2}$~\cite{bosonization}, respectively,
with $\tilde{V}(0)=A^{1-\beta}/(\beta-1)$ the Fourier transform of $V(y)$ at zero-momentum, and $\zeta$ the Riemann Zeta-function. Due to the similar functional dependence, we then fix $A=[\beta(\beta-1)(\beta+1)\zeta(\beta)/\pi]^{1/(1-\beta)}$ by matching $K_{\rm w}=K_{\rm s}$ for $n^{\beta-2}R_{\beta}\gg 1$ and obtain the approximate Eq.~\eqref{eq:eqK}~\footnote{This fails for the Coulomb case~\cite{schulz} [$\zeta({\beta})$ not defined], with the cut-off usually fixed by the microscopic theory.}.


Expression~\eqref{eq:eqK} compares favorably with known exact results.
In Fig.~\ref{fig:fig3}(a) (Inset) we compare it to the exact expression $K_{CS}=2/(1+\sqrt{1+2R_{2}})$ for $\beta=2$, which we derive from the Bethe-Ansatz solution of
the CS-model~\cite{calogero1}. We find \textit{quantitative} agreement
between the two curves for the entire range of interaction strengths
$0< R_{2}\leq100$, with a maximal relative difference of about
5\% at $R_{2}\simeq1$, and recover the $n$-independence of the CS model~\cite{calogero1}. Furthermore, in the main figure we compare
$K=1/\sqrt{1+0.73nR_{3}}$, as derived from Eq.~\eqref{eq:eqK} for
$\beta=3$, to the numerical
quantum Monte-Carlo results of Refs.~\cite{citro} and~\cite{roscilde}
(black squares and red dots, respectively), finding good
agreement for $0<nR_{3}\leq1000$. In panel
(b), we also plot the velocity $v$ 
in the same range of $nR_{3}$ values, finding excellent agreement with the results
of~\cite{citro}. This fixes the phenomenological parameters
in the effective Hamiltonian~\eqref{eq:eqLL}. We are not aware
of exact results for $\beta>3$ to compare with our predictions.

The extension of the techniques described here to
several species will enable a microscopic treatment
of strongly correlated phenomena in mixtures
of polar molecules in single- and multi-tube configurations,
as relevant to experiments~\cite{Ni}, in particular exotic phases such as bond-ordered density waves and trimer liquids.

We thank M. Di Dio, E. Ercolessi, R. Fazio, F.~Ferlaino and J. Ye for discussions, H.J. Kimble and P. Julienne for  hospitality at Caltech and JQI. This work was supported by U.Md. PFC/JQI, MURI, EOARD FA8655-10-1-3081, the Austrian FWF, the EU through NAME-QUAM.

\end{document}